\documentclass[preprint2]{aastex}

\shortauthors{Gizis et al.}
\shorttitle{No Brown Dwarf Desert}

\begin{document}

\title{Substellar Companions to Main Sequence Stars:  
No Brown Dwarf Desert at Wide Separations}

\author{John E. Gizis\altaffilmark{1},
 J. Davy Kirkpatrick\altaffilmark{1},
 Adam Burgasser\altaffilmark{2},
 I. Neill Reid\altaffilmark{3},
 David G. Monet\altaffilmark{4},
 James Liebert\altaffilmark{5} and
 John C. Wilson\altaffilmark{6}}

\altaffiltext{1}{Infrared Processing and Analysis Center, 100-22, 
  California Institute of Technology,
  Pasadena, CA 91125 \email{gizis@ipac.caltech.edu}}
\altaffiltext{2}{Department of Physics, 103-33, California Institute of 
Technology, Pasadena, CA 91125}
\altaffiltext{3}{Department of  Physics and Astronomy, 
University of Pennsylvania, 209 South 33rd Street, Philadelphia, PA 19104-6396}
\altaffiltext{4}{U.S. Naval Observatory, P.O. Box 1149, Flagstaff, AZ 86002}
\altaffiltext{5}{Steward Observatory, University of Arizona, Tucson, AZ 85721}
\altaffiltext{6}{Space Sciences Building, Cornell University, Ithaca, NY 14853}

\begin{abstract}
We use three field L and T dwarfs which were discovered to be 
wide companions to known stars by the Two Micron
All-Sky Survey (2MASS) to derive a preliminary
brown dwarf companion frequency.  Observed L and T dwarfs
indicate that brown dwarfs are not unusually rare as
wide ($\Delta >1000$ A.U.) systems to F-M0 main-sequence
stars ($M>0.5M_\odot$, $M_V<9.5$), even though they are rare
at close separation ($\Delta <3$ A.U.), the ``brown dwarf desert.''  
Stellar companions in these separation ranges are equally frequent, but
brown dwarfs are
$\gtrsim 10$ times as frequent for wide than close separations.
A brown dwarf wide-companion frequency as low as
the $0.5$\% seen in the brown dwarf desert is ruled out
by currently-available observations.  
\end{abstract}


\section{Introduction}

Understanding the processes, and distinctions between, 
star formation, binary formation and planetary formation
is a major goal of stellar astronomy. A necessary step 
towards that goal is
to understand the frequency of ``brown dwarf'' and ``planetary''
companions as a function of separation, primary mass, and
secondary mass.  It is now well-known that brown dwarfs
are very rare as close companions to F-M dwarfs (the ``brown dwarf
desert,'' Marcy \& Butler 2000) yet they are quite common in the 
field \citep{reidmf} and in open clusters \citep{bsmbwb98,martin98}.  

The purpose of this letter is to show that current data
allow an estimate of the wide ($\Delta >1000$ A.U.) brown dwarf companion
frequency to near-solar mass main sequence stars, despite the present 
lack of well-defined
searches for wide companions.  The observational constraints and resulting 
wide companion frequency are discussed in
Section~\ref{obs} and the differences between this fraction
and the close companion frequency are discussed in Section~\ref{discussion}.

\section{Observational Constraints\label{obs}}

2MASS searches for ultracool dwarfs are based on 
colors and magnitudes; although designed to find
isolated field dwarfs, they are not biased against
widely separated ($\gtrsim 40$ arcsec) companions 
whose photometry is uncontaminated by the primary star. 
Three published brown dwarf companions with separations $\Delta>1000$ A.U. 
have been identified in the course of
2MASS searches for isolated field brown dwarfs.
The L4.5 dwarf Gl 417B has an estimated mass of $0.035\pm0.015M_\odot$ and
age $0.08-0.3$ Gyr while  the L8 dwarf 
Gl 584C has a mass of $0.060 \pm 0.015 M_\odot$ and an age of 1.0-2.5 Gyr
\citep{k00,k01}.
The T dwarf Gl 570D has a mass of $0.050 \pm 0.020 M_\odot$ and an
age of 2-10 Gyr \citep{gl570d}.  All are definitely substellar 
via spectroscopic crieria as well, 
since both L dwarfs pass the lithium test while the 
T dwarf is too cool to be a star.
All three have been confirmed as companions by their common 
proper motion.  These reflect the results from 2MASS searches of 
only a fraction of
the sky --- wide brown dwarf companions continued to be 
discovered.  \citet{gl337c} will describe the discovery of 
three additional L dwarf secondaries
at 880 A.U., 1090 A.U., and 2460 A.U.  

The usual procedure for determining companion frequency
is to survey a number ($N$) of potential primary stars. 
If $n$ brown dwarfs with spectral type L 
are found, the frequency of L dwarf companions (in the searched
separation range) is simply $f_L = \frac{n}{N}$.  In the case
of imaging surveys, only a fraction ($y_L$) of the brown dwarfs
can be detected as L dwarfs, since brown dwarfs fade to 
very low temperatures and luminosities.  Modelling of $y_L$ allows an
estimate of the true brown dwarf frequency to be $f_{bd} = \frac{n}{y_LN}$.  
In principle, the Two Micron All-Sky Survey (2MASS)  
allows a search for wide companions
to nearby stars, allowing $f_{L}$ and $f_{bd}$ to be determined.  A complete
search, however, has not yet been made, due to the
complexities of the task. 

It is nevertheless possible to estimate the frequency of 
wide-separation brown dwarf companions, since the frequency
can also be expressed as 
$$f_{bd} = \frac{\rho_{comp}}{\rho_{star}} = 
\frac{\rho_{comp}}{\rho_{bd}}\frac{\rho_{bd}}{\rho_{star}} =
\frac{g_{w}\rho_{bd}}{\rho_{star}} = 
\frac{g_{w}\rho_{L}}{y_{L}\rho_{star}}
$$  In this equation, $\rho_{star}$ is the space density of 
stars (the potential primaries), 
$\rho_{comp}$ is the space density of wide brown dwarf
companions, $\rho_{bd}$ is the
space density of field brown dwarfs found by 2MASS, 
and $g_w$ is the frequency of field brown dwarfs that have a stellar
primary at $\Delta >1000$ A.U.  Each of these variables can be estimated
from published data. 
Separations of 1000 A.U. correspond to 
40 arcseconds at 25 parsecs distance, and larger values
at closer distances.  For these wide systems, the presence
of the main sequence star does not affect the identification 
of the brown dwarf in searches for isolated field objects.  
In the final step, we have assumed that $g_{w}$ for all
brown dwarfs is the same as $g_{w}$ for L dwarfs.
\footnote{Some of the field L dwarfs used in our
calculation will not be brown dwarfs; however, since
both Gl 417B and Gl 584C are confirmed brown dwarfs, accounting
for this effect will only {\it increase} the derived brown dwarf
companion frequency.}

The observed space density of L dwarfs can be determined
using the \citet{k99} sample of L dwarfs.  Adopting the
limiting magnitude $K_s=14.7$ and the \citet{k00} 
spectral type-$M_{K_s}$ relation, the \citet{s68} 
${1}\over{V_{max}}$ technique yields an L dwarf space density of
$\rho_L = 0.0057 \pm 0.0025$ pc$^{-3}$.  The space density of T dwarfs 
is highly uncertain but comparable \citep{b99}.  Only a small
fraction ($y_{L}$) of brown dwarfs are observable as L, or even T, 
dwarfs.  We can estimate $y_{L}$ in two ways.  
{\it If} the mass function of companions is similar to that 
of isolated field brown dwarfs, then we can use the correction
determined for isolated brown dwarfs.  
\citet{reidmf} estimate that {\it if} the substellar
mass function can be described as a power law
$\frac{dN}{dM} \propto M^{-\alpha}$ extending down
to $0.01 M_\odot$, then 
$\alpha \approx 1.3$ and $\rho_{bd} \approx 0.10$ b.d.  pc$^{-3}$ 
with large uncertainties.  For the cases $\alpha=0.0,~0.6$ and $1.0$,
then $\rho_{bd} \approx 0.02,~0.04$ and 0.07 b.d.  pc$^{-3}$ 
respectively. Surveys of the Pleiades are consistent with 
$\alpha \approx 0.6-1.0$ \citep{bsmbwb98,martin98}, and neither
that cluster nor the field is consistent with $\alpha=0$.
We adopt $\rho_{bd} = 0.07 \pm 0.03$ b.d.  pc$^{-3}$; hence, the 
ratio of L dwarfs to brown dwarfs is $y_L \approx 0.08$.   

Alternatively, we may attempt to estimate the parameter 
$y_{L}$ directly from the properties of the L dwarf companions.
Gl 584C is just at the limits of detectability; since its
age is in the range 1.0-2.5 Gyr, it is only detectable
for 0.1-0.25 of the age of the disk.  Gl 417B is easily detectable
as an L4.5, but would have been detectable down to spectral type
L8.  Comparison to evolutionary tracks indicates it would then
be visible to an age of 0.3-1.0 Gyr.
The companions themselves then imply a correction
factor $y_L$ in the range 0.07 - 0.18 --- consistent with
the isolated brown dwarf estimate.  

The fraction of the ``isolated'' field brown dwarfs which
are actually wide companions to main sequence stars
can be estimated from the 2MASS field brown dwarf searches.  
Over 100 field L dwarfs are now published, primarily
from the 2MASS \citep{k99,k00}, DENIS \citep{delfosse},
and SDSS \citep{fan} surveys, of which two are wide companions to 
near-solar mass stars.  Many of the L dwarfs, however, are at large distances
($>25$ pc), where it would be difficult to discover an L dwarf 
companion due to proximity of the 
primary star.  We choose to consider only L dwarfs within
25 parsecs because at this distance, we can be confident that bright (F-K) 
primaries are cataloged and the L dwarf companions at $\Delta >1000$ A.U. 
will be detectable.   In this case, 2 of 40 L dwarfs \citep[Table 4]{k00}
have primaries, indicating the fraction of field L dwarfs which are
actually wide companions is $g_w = 0.05 \pm  0.04$.
This estimate is supported by the ongoing 2MASS T dwarf search
\citep{b99}, which has now discovered 17 field T dwarfs
\cite{b_conf}.  Since T
dwarfs are simply the older counterparts of L dwarfs, the
fraction of 'field' dwarfs which are companions
should be roughly the same.  One (Gl 570D) of the
T dwarfs is the fourth
member of a K/M/M/T quadruple system \citep{gl570d}, leading 
to an estimate of $g_w = 0.06 \pm 0.06$.  (Note that Gl 229B is too close
to Gl 229A to be counted or even detected by 2MASS.)  
Combining the L dwarf and T dwarf values, we adopt $g_w = 0.05 \pm 0.03$.  

The space density of stars which are potential primaries for the
brown dwarfs can be determined from the 
nearby star catalog as modified by Hipparcos parallaxes.  
All of the primaries to wide L/T dwarf companions 
(including the new Wilson et al. 2001 systems),have $M_V<9.5$
(or Mass $> 0.5 M_\odot$).  
\citet{jw97}
find that the space density of stars with $M_V < 9.5$ 
is $0.020 \pm 0.001$ pc$^{-3}$.  These are 
F to M0 main sequence stars.  We do not consider fainter
primaries because the incompleteness of 
the late-M dwarf catalog means that we might not 
recognize their presence near an L dwarf.  We therefore have
insufficient information to calculate the M/L dwarf binary statistics.
\footnote{Note that the early-M dwarfs with $9.5 \le M_V < 13.5$ 
contribute another 
$0.039 \pm 0.008$ stars pc$^{-3}$, implying that we could 
decrease the derived brown dwarf frequency by a factor 2.9 --- but in this
case we face the puzzling fact that the chance of all 5 primaries
having $M_V<9.5$ is only $\sim 0.4$\%.  
An analogous difficulty in interpretation would occur if 
a search for companions amongst F-M dwarfs found companions
only around the F, G and K dwarfs.}
We can now estimate the wide brown dwarf companion frequency
to main sequence stars.  Taking the stars with $M_V < 9.5$ 
as the available primaries, we find $f_L = 0.014 \pm 0.011$. 
These, however, represent only a small fraction of the total
brown dwarf population.  We estimate 
$f_{bd} = 0.18 \pm 0.14$ by assuming a mass function 
$\alpha = 0.7$; the $y_L$ values from the companions themselves
suggest values in the range 0.08-0.20 $\pm 0.14$.  

\begin{figure}
\epsscale{0.7}
\plotone{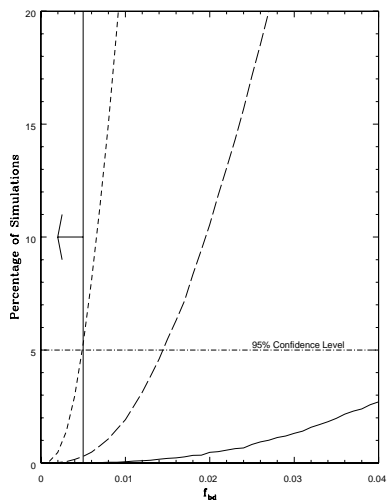}
\caption{The percentage of samples in which 3 wide companions are
found in a sample of 57 field brown dwarfs.  The solid curve
is our preferred model ($\rho_{bd}=0.07$,$\rho_{star}=0.020$); 
the long-dashed line is a model with few brown dwarfs
($\rho_{bd}=0.02$,$\rho_{star}=0.020$); the short-dashed curve
is a model with few brown dwarfs and M dwarf primaries
($\rho_{bd}=0.02$,$\rho_{star}=0.059$).  The position of the
brown dwarf desert ($f_{bd} \le 0.005$) and the 95\% 
confidence level is also marked.  
\label{fig_prob}
}
\end{figure}

The large uncertainties --- due to the fact that the estimates
are based only 3 companions --- indicate that a larger survey is
needed to determine the brown dwarf companion fraction; we
certainly cannot prefer $f_{bd}=0.18$ over 
$f_{bd}=0.05$ given the uncertainties.  
However, it is important to realize that the error bars are
non-Gaussian and that very low companion fractions ($f_{bd}$) are
already ruled out with statistical significance.  
The situation is illustrated in Figure~\ref{fig_prob}.  
We run Monte Carlo simulations of samples of 57 brown dwarfs
and determine the percentage that have at least 3 
wide companions.  (For each value of $f_{bd}$, we run
200,000 simulations.)  The solid line plots our preferred
scenario:  A brown dwarf space density of 0.07 pc$^{-3}$
(equivalently, $y_L = 0.08$) and all primaries with
$M_V<9.5$ considered.  It is possible to consider other
scenarios.  If $\rho_{bd} = 0.02$, equivalent to assuming
$\alpha=0$ or $y_L = 0.29$ (long-dashed line), then values of
$f_{bd}$ as low as 0.015 can still be excluded 
at the 95\% confidence level.  
This corresponds to a scenario in which  
nearly all brown dwarfs are relatively massive 
and therefore detectable as L dwarfs for many Gyr; in this case,
the observed L dwarfs represent a larger fraction of
the total brown dwarf population.  We finally consider 
an extreme model, in which $\rho_{bd} = 0.02$
and $\rho_{star} = 0.059$ pc$^{-3}$.  In this case, the brown dwarf
binary fraction can be much lower; in other words,
we can reduce the reported brown dwarf fraction by averaging
the many brown dwarf companions to $M>0.5M_\odot$ primaries
with the more numerous population of lower mass stars
for which no wide brown dwarf companions have ever been
detected.

\section{Discussion\label{discussion}}

The existence of the ``brown dwarf
desert'' separating stars from ``planets'' at separations
less than 3 A.U. is well established
\citep{hippbd}.    
The review of \citet{mb00} finds that {\it less than} 0.5\% of 
F-M dwarfs have brown dwarf companions down to $0.01 M_\odot$
on the basis of over 500 stars.  
The \citet{mazeh} analysis of
the 164 nearby G dwarfs studied by \citet{dm91} indicates
that $13\pm 3$\% of G dwarfs have {\it stellar} companions 
in this separation range, and that 
the mass distribution of these close companions is lacking in very-low-mass
stars (a ``red dwarf steppe''?) compared to both wider 
companions and the field mass function.
On the basis of \citet{dm91}'s analysis (their Figure 7), we estimate that 
$12 \pm 3$ \% of G dwarfs have stellar companions with separation
$\Delta >1000$ A.U.  

Our evidence that F-M0 dwarfs have a significant
($f_{bd}=18\pm14$\%) population of very-low-mass companions
at wide separations indicates that the 
orbital separation distribution of brown dwarf  
companions is not simply a scaled-down version of the
stellar companion distribution.  
G dwarfs have approximately equal numbers of 
stellar-mass companions at separations $\Delta <3$ A.U. 
and $\Delta > 1000$ A.U.  ($r_{1000/3} \approx 0.9$);
in contrast, we estimate their 
brown dwarf companions are at least four, and probably many more, 
times more common
at the larger separations ($r_{1000/3} \gtrsim 4$).  
Within the very large uncertainties, the 
number ratio of stellar and brown dwarf companions may be 
near unity, similar to the ratio for isolated stars and brown dwarfs
\citep{reidmf}, although a considerably larger sample is
needed to investigate this.  
The fact that all the primaries 
are relatively massive suggests that wide brown dwarf
companions do not form around --- or are not retained by ---
less massive ($M<0.5M_\odot$) primaries, although the
incompleteness of the nearby star catalog for 
M dwarfs may contribute to this effect.  The extreme
limit to this is noted by \citet{reidhst}, who find
that L dwarf primaries lack companions beyond $\sim 10$ A.U.  
Where the ``brown dwarf desert'' for F-M0 dwarf primaries ends
is unclear at present, but searches around stars in the range 1-100 A.U. 
have generally had little success even though this range is the 
peak of stellar companion distribution for both G and M dwarf primaries
\citep{dm91,fm92}.  Most recently, \citet{shst} failed to 
discover any brown dwarfs despite a sensitive 
{\it Hubble Space Telescope} search 
at separations of $1-60$ A.U. around 23 stars, while the
extensive, highly sensitive search of 107 stars 
by \citet{bro2001} found only 1 brown
dwarf in range $40-120$ A.U.  (Note that these volume-limited samples
are dominated by M dwarf primaries with $M<0.5M_\odot$.)  The fraction in
the range 100-1000 A.U. cannot yet be constrained, 
but the recent discoveries of 
three L brown dwarf companions at those separations \citep{g1963b,lhs102b,g01}
suggest that they {\it may} be as common as wide companions
and that 2MASS should detect many more.

While the uncertainty of our derived companion fraction
is large, perspective on the brown dwarf desert 
may be gained by considering the numbers of stars and brown dwarfs within
25 parsecs.  There are 1297 main sequence stars with $M_V<9.5$
in this volume according to the \citet{jw97} luminosity function.
Radial velocity surveys indicate that they will have fewer
than 6.5 brown dwarf companions within 3 A.U.  In constrast, 
three wide-separation (two L, one T) brown dwarf companions
are already known to such stars
even though 1) only a small fraction of the sky has been searched,
2) T dwarfs are not detectable all the way to 25 parsecs,
and 3) older/lower-mass brown dwarfs are undetectable.  
Our numbers suggest another
$\sim 16\pm9$ wide L dwarf companions within 25 parsecs will be found,
representing only the young, massive tip of the brown dwarf population.  

Combined with the \citet{reidhst} observation that 
20\% of L dwarfs have H.S.T. resolved ($>0.1$ arcsec) 
brown dwarf companions
within $10$ A.U. but that wider companions are lacking, 
(see also Koerner et al. 1999), there is strong evidence 
that the frequency of brown dwarf
companions is strongly dependent on both primary mass
and orbital separation.  There is a ``brown dwarf
desert'' at $<3$ (or $\lesssim 100$?) A.U. for 
F to mid-M main sequence primaries,
and another desert for wide ($\gtrsim 20$ A.U.) brown dwarf doubles.  
This situation arises naturally in some theories of star formation
\citep{boss,bate}.

\section{Summary\label{summary}}

We use the field L and T dwarfs which were discovered to be 
wide companions to known stars to derive a preliminary 
brown dwarf companion frequency.  The observed L and T dwarfs
indicate that brown dwarfs are not unusually rare as
wide ($>1000$ A.U.) systems, even though they are rare ($<0.5\%$)
at close ($<3$ A.U.)separations.  The current data indicate that
$\sim 1\%$ of $M_V<9.5$ primaries have a wide L dwarf companion; 
the brown dwarf fraction should be substantially ($5-13$ times) 
higher.  Stellar companions in these separation ranges are equally frequent.  

Our estimates indicate that continued searches for wide brown 
dwarf companions should not be discouraged by the 
existence of the ``brown dwarf desert'' at close
separations.  There is now strong evidence --- from both
the wide companions examined in this work, and the close
double brown dwarfs examined by \citet{reidhst} --- that 
the brown dwarf companion frequency is a strong function
of both primary mass and separation.  
A search using 2MASS for both stellar and brown dwarf companions 
with separation $\Delta >100$ A.U. will be quite rewarding.  Furthermore,
provided the older and lower-mass wide brown dwarf companions 
are not stripped in the Galactic disk, the majority of the brown 
dwarfs should be intrinsically cooler 
than Gl 570D, suggesting that a deep wide-field SIRTF search near
G and K dwarfs is a promising avenue to extend our knowledge
of brown dwarfs.  

\acknowledgments

J.E.G. and J.D.K. acknowledge the
support of the Jet Propulsion Laboratory, California
Institute of Technology, which is operated under contract
with NASA.  
This work was supported by NASA's Jet Propulsion 
Lab through contract number 961040NSF, a core science grant to the 
2MASS science team.
This publication makes use of data products from 2MASS, 
which is a joint project of the
University of Massachusetts and IPAC/Caltech, funded by NASA and NSF.

%Tables look like this

%--------------------------BIBLIOGRAPHY---------------------------

\end{document}